\documentclass[aps,prd,twocolumn,superscriptaddress,showpacs,nofootinbib,nobibnotes]{revtex4}
\usepackage[dvips]{epsfig}
\usepackage{amsmath}

\begin{document}

\title{Tau Neutrino Astronomy in GeV Energies}

\author{H. Athar}
\email{ athar@phys.cts.nthu.edu.tw}
\affiliation{Physics Division, National Center for Theoretical Sciences, Hsinchu 300, Taiwan}
\affiliation{Institute of Physics, National Chiao-Tung University, Hsinchu 300, Taiwan}

\author{Fei-Fan Lee}
\email{ u1717117.py87g@nctu.edu.tw}

\author{Guey-Lin Lin}
\email{ glin@cc.nctu.edu.tw}
\affiliation{Institute of Physics, National Chiao-Tung University, Hsinchu 300, Taiwan}
\date{\today}
\begin{abstract}
We point out the opportunity of the tau neutrino astronomy for the neutrino
energy $E$ ranging between  10 GeV and $10^{3}$ GeV. 
In this energy range, the intrinsic tau neutrino production is
suppressed relative to the intrinsic muon neutrino production. Any 
sizable tau neutrino flux may thus arise because of the  
 $\nu_{\mu}\to \nu_{\tau}$ neutrino oscillations only. It
is demonstrated that, in the presence of the neutrino oscillations, 
 consideration of the neutrino flavor dependence in the background
atmospheric neutrino flux leads to the drastically different prospects
between the observation of the astrophysical muon neutrinos and that of the
astrophysical tau neutrinos. 
 Taking the galactic-plane neutrino flux
as the targeted astrophysical source, we have found that the
galactic-plane tau neutrino flux dominates over the atmospheric tau
neutrino flux for $E \geq $ 10 GeV. Hence, the
galactic-plane can at least in principle be seen through the tau neutrinos with
energies just greater than 10 GeV. In a sharp contrast, the
galactic-plane muon neutrino flux is overwhelmed by its atmospheric
background until $E \geq 10^{6}$ GeV.
\end{abstract}
\pacs{98.38.-j, 13.85.Tp, 14.60.Pq}
\maketitle

\section{Introduction} 

The $\nu_{\mu}\to \nu_{\tau}$  neutrino 
 oscillations established by the high-statistics
Super-Kamiokande (SK) detector, ensure that a non-negligible 
 $\nu_{\tau}$ flux reaches the Earth. A recent SK analysis of the
atmospheric neutrino data implies the following range 
 of the neutrino mixing parameters \cite{Ashie:2004mr}
\begin{equation}
 \delta m^{2}=(1.9-3.0)\cdot 10^{-3}\, \, \, {\rm eV}^{2}, \, \, \,
 \sin^{2}2\theta >0.9.
\label{range}
\end{equation}
This is a  $90\% \, {\rm C.L.}$ range 
 of the neutrino mixing parameters with the best fit values given
by $\delta m^{2}=2.4\cdot 10^{-3}\, \, \, {\rm eV}^{2}$ and
$\sin^{2}2\theta =1$ respectively.

The tau neutrinos resulting from the above
$\nu_{\mu}\to \nu_{\tau}$ oscillations are presently
  identified on the statistical basis \cite{Suzuki:2003hn}. 
 On the other hand, the total number of observed
non-tau neutrinos from various detectors are already greater than
$\sim 10^{4}$ with energies ranging from $\sim 10^{-1}$ GeV 
 to $\sim 10^{3}$ GeV
\cite{Kajita:2000mr}. It is essential to develop efficient
techniques to identify the tau neutrinos \cite{Athar:2002rr}.

There are at least two important reasons for observing 
 the $\nu_{\tau}$. First,
seeing the $\nu_{\tau}$ confirms the $\nu_{\mu}\to \nu_{\tau}$
oscillation interpretation for the atmospheric neutrino data.
Second, since the atmospheric $\nu_{\tau}$ flux is generally
suppressed as compared to the atmospheric $\nu_{\mu}$ flux, the
prospective observation of the astrophysical $\nu_{\tau}$ suffers 
 {\tt much less}
background than in the $\nu_{\mu}$ case. 

In this paper, we
address the second point with the galactic-plane tau neutrinos as our
illustrating astrophysical source. 
We point out that {\tt contrary} to the general expectations, the
atmospheric neutrino flux {\tt does not} dominate 
 over the astrophysical
neutrino flux for the neutrino energy $E< 10^{3}$ GeV, once the
flavor composition of the  neutrino flux is
taken into account. The idea for such an
investigation has appeared earlier in Ref.~\cite{Athar:2004pb}.

In the context of two neutrino flavors, $\nu_{\mu}$ and
$\nu_{\tau}$, the total tau neutrino flux arriving at the detector
 on Earth, after traversing  a distance $L$, is
\begin{equation}
 \phi_{\nu_{\tau}}^{\rm tot}(E)=P(E)\cdot
 \phi_{\nu_{\mu}}(E)+(1-P(E))\cdot \phi_{\nu_{\tau}}(E),
\label{osc}
\end{equation}
where $P(E)\equiv P(\nu_{\mu}\to \nu_{\tau})=\sin^2 2\theta\cdot
\sin^2(L/L_{\rm osc})$ with the neutrino oscillation length given by
$L_{\rm osc}=4E/\delta m^2$.

In order to compute the total $\nu_{\tau}$ flux from a given
astrophysical site, we need to first compute the intrinsic $\nu_{\mu}$ 
as well as the intrinsic $\nu_{\tau}$ flux from the same site. 

\section{Galactic-plane neutrino flux}

 One calculates the intrinsic
galactic-plane $\nu_{\mu}$ and $\nu_{\tau}$ fluxes by considering
the collisions of incident cosmic-ray protons with the interstellar
medium. The fluxes are given by
\begin{equation}
 \phi_{\nu}(E)= Rn_{p}\int_{E}^{\infty}
 \mbox{d}E_p \; \phi_{p}(E_{p}) \, 
 \frac{\mbox{d}\sigma_{pp \to \nu +Y}}{\mbox{d}E},
\label{gala}
\end{equation}
where $E_p$ is the energy of the incident cosmic-ray proton,
${\rm d}\sigma_{pp\to \nu+Y}/{\rm d}E$ is the $\nu $ energy 
 spectrum  in the $pp$ collisions. $R$ is the typical
\begin{figure}[t]
\begin{center}
\epsfig{file=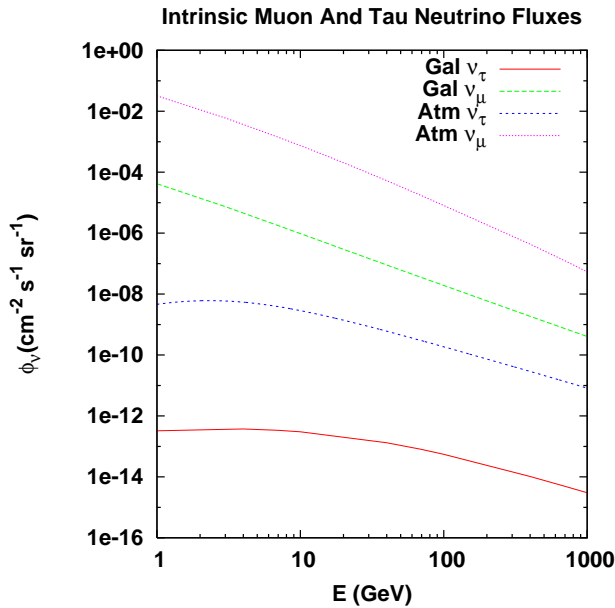,width=3.25in}
\caption{The intrinsic galactic-plane and the intrinsic 
 downward going atmospheric
 muon and tau neutrino fluxes
 as a function of the neutrino energy in GeV.
 Details are provided in the text.}
\label{Fig1}
\end{center}
\end{figure}
distance in the galaxy along the galactic-plane, which we take as
$10$ kpc (1 pc $\sim 3\cdot 10^{13}$ km). The density
of the interstellar medium $n_{p}$ along the galactic plane 
 is taken to be $\sim $ 1 proton per
cm$^3$. The primary cosmic-ray proton flux,
$\phi_p(E_p)\equiv {\rm d}N_p/{\rm d}E_p$, 
 is given by \cite{Gaisser:2002jj}
\begin{equation}
 \phi_p(E_{p})=1.49\cdot \left(E_{p}+2.15\cdot
 \exp(-0.21\sqrt{E_{p}})\right)^{-2.74},
\label{cosmic}
\end{equation}
in units of cm$^{-2}$s$^{-1}$sr$^{-1}$GeV$^{-1}$. 
The above flux is under the assumption that the cosmic-ray
flux spectrum in the galaxy is a constant and equal to its locally
observed value. The galactic-plane neutrino flux 
 (abbreviated here as Gal) is  
sometimes also referred to as the galactic center region 
  neutrino flux (abbreviated as G or GC), 
 the galactic disk neutrino flux or the
Milky Way neutrino flux. We shall estimate here the 
neutrino flux coming from the galactic-plane direction only 
 as transverse to it, the $n_{p}$
decreases essentially exponentially \cite{Ingelman:1996md},
and so does the $\phi_{\nu}(E)$
according to Eq. (\ref{gala}).

The neutrino production process $p+p\to
\nu+Y$ is mediated by the production and the decays of 
 the $\pi$, the $K$, and
the charmed hadrons. The galactic muon neutrinos mainly come from the 
 two-body $\pi$
decays and the subsequent three-body muon decays. While the decay
rates and the decay distributions of the $\pi$ and the $\mu$ are well
understood, the differential cross 
 section for the process $p+p\to \pi+Y$ is
model dependent. We adopt the parameterization  for this
 cross section from Ref. \cite{Gaisser:2001sd}, which is obtained by 
 using the accelerator data in
the sub-TeV energy range \cite{Gaisser:1990vg}. We remark that 
 our galactic-plane
$\nu_{\mu}$ flux compares well with a previous calculation
\cite{Ingelman:1996md} using the PYTHIA \cite{Sjostrand:1993yb}. 

The galactic-plane $\nu_{\tau}$
  flux arises from the production and the decays of the $D_s$ mesons. It
has been found to be rather suppressed compared to the corresponding
 $\nu_{\mu}$ flux \cite{Athar:2001jw}. 
Fig. \ref{Fig1} shows the intrinsic galactic-plane $\nu_{\mu}$  and 
 $\nu_{\tau}$ fluxes obtained by using the above description 
 (along with the corresponding intrinsic downward 
 going atmospheric neutrino
 fluxes). We shall 
use these in our subsequent estimates.

The total galactic-plane
tau neutrino flux, $\phi_{\nu_{\tau}}^{\rm tot}(E)$, is therefore 
 {\tt dominated} by
the $\nu_{\mu}\to \nu_{\tau}$ oscillations indicated by the term
$P(E)\cdot \phi_{\nu_{\mu}}(E)$ in Eq.~(\ref{osc}). With the
best-fit values for the neutrino mixing parameters,
 we have $\phi_{\nu_{\tau}}^{\rm tot}(E)\approx
\phi_{\nu_{\mu}}(E)/2$, neglecting the contribution of
$\phi_{\nu_{\tau}}(E)$, since $L_{\rm osc} \ll L$, where $L \sim $ 5 kpc. 
 The {\tt total} galactic-plane tau neutrino 
 flux, $\phi^{\rm tot}_{\nu_{\tau}}(E)
 \equiv {\rm d}N/{\rm d}({\rm
log}_{10}E)$,  can be parameterized  for 1 
 GeV $\leq E\leq 10^{3}$ GeV, as

\begin{equation}
 \phi^{\rm tot}_{\nu_{\tau}}(E)=A \left(\frac{E}{\rm GeV}\right)^{\alpha}, 
\label{gal-fit}
\end{equation}
where $A = 2\cdot 10^{-5}$  is in units 
of cm$^{-2}$s$^{-1}$sr$^{-1}$ with 
 $\alpha = -1.64$. 
\section{Atmospheric neutrino flux} 

The calculation of the total
atmospheric tau neutrino flux is more involved. We follow the
approach in Ref. \cite{Gaisser:2001sd} for 
 computing the intrinsic
atmospheric $\nu_{\mu}$ flux which can oscillate into $\nu_{\tau}$.

For the $\pi$-decay contribution, the flux formula reads:
\begin{eqnarray}
 \frac{\mbox{d}^2N^{\pi}_{\nu_{\mu}}(E,\xi,X)}{\mbox{d}E\mbox{d}X}&=&
 \int^{\infty }_{E}
 \mbox{d}E_N\int^{E_{N}}_{E}
 \mbox{d}E_{\pi}\frac{\Theta(E_{\pi}-\frac{E}{1-\gamma_{\pi}})}
 {d_{\pi}E_{\pi}(1-\gamma_{\pi})}
 \nonumber \\
 & &\times \int_{0}^{X}
 \frac{\mbox{d}X'}{\lambda_N}P_{\pi}(E_{\pi},X,X')\nonumber \\
 & &\times \frac{1}{E_{\pi}}F_{N\pi}(E_{\pi},E_N)\nonumber \\
 & & \times \exp \left(-\frac{X'}{\Lambda_N}\right)\phi_N(E_N),
\label{atm-nu}
\end{eqnarray}
where $E$ is the neutrino energy and $\xi$ is the zenith angle in the
direction of the incident cosmic-ray nucleons. The
$\gamma_{\pi}=m_{\mu}^2/m_{\pi}^2$ and $d_{\pi}$ is the pion decay
length in units of g/cm$^2$. The $\lambda_N$ is the nucleon interaction
length while $\Lambda_N$ is the corresponding nucleon attenuation
 length. $\phi_N(E_N)$ is the primary cosmic-ray spectrum. We
only consider the proton component of $\phi_N$, which is given by
Eq.~(\ref{cosmic}). We take $\lambda_N=86$ g/cm$^2$ and
$\Lambda_N=120$ g/cm$^2$ \cite{Gaisser:1990vg}. The function
$P_{\pi}(E_{\pi},X,X')$ is the probability that a charged pion
produced at the slant depth $X'$ (g/cm$^2$) 
 survives to the slant depth
$X$ ($> X'$). The $F_{N\pi}(E_{\pi},E_N)$ is 
 the normalized inclusive cross
section for $N+{\rm air}\to \pi^{\pm}+Y$, and is  
 given in the Ref.~\cite{Gaisser:2001sd}.

The kaon contribution to the atmospheric $\nu_{\mu}$ flux has the
same form as Eq.~(\ref{atm-nu}) with an inclusion of the branching
ratio $B(K\to \mu\nu)=0.635$ and appropriate replacements in the
kinematics factors as well as in the normalized inclusive cross section. 
 We
remark that the current approach neglects the 3-body muon-decay
contribution to the $\nu_{\mu}$ flux. This is a good approximation
for $E > 10$ GeV \cite{Gaisser:2001sd}. The relevance of 
 the muon-decay contribution
for 1 GeV $\leq E \leq 10$ GeV will be commented later.

We stress that the $\pi$ and the $K$ decays are not 
 the only sources for the 
\begin{figure}
\epsfig{file=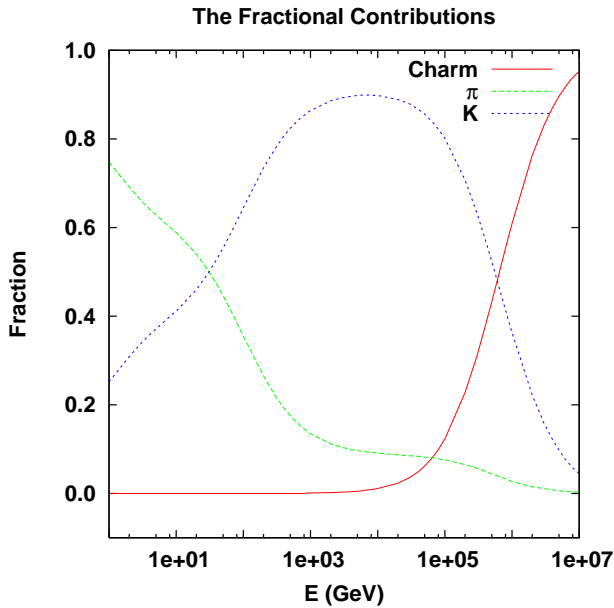,width=3.25in}
\caption{The fraction of contributions by the $\pi$, the 
$K$, and the charm decays to the overall downward going atmospheric 
 $\nu_{\mu}$ flux as a function of the neutrino energy in GeV.}
\label{Fig2}
\end{figure}
atmospheric $\nu_{\mu}$ flux. For $E > 10^6$ GeV, the charm-decay
contribution becomes more important than those of the $\pi$ and the $K$
decays. We have used the results from the perturbative QCD
 to estimate this contribution \cite{Pasquali:1998ji}. 
 The muon neutrino flux 
due to charm contribution can be written as
\begin{equation}
 \frac{\mbox{d}^2N^c_{\nu_{\mu}}(E,X)}{\mbox{d}E\mbox{d}X}=\sum_h
 \frac{Z_{ph}Z_{h\nu_{\mu}}}{1-Z_{pp}(E)}\cdot
 \frac{\exp(-X/\Lambda_p)\phi_p(E)}{\Lambda_p},
\label{mu-charm}
\end{equation}
where $h$ stands for the $D^{\pm}$, the $D^{0}$, the $D_{s}$ 
 and the $\Lambda_{c}$
hadrons. The $Z$ moments on the RHS of the equation are defined
by
\begin{equation}
 Z_{ij}(E_j)\equiv \int_{E_j}^{\infty}{\mbox
 d}E_i\frac{\phi_i(E_i)}{\phi_i(E_j)}\frac{\lambda_i(E_j)}{\lambda_i(E_i)}
 \frac{{\mbox d}n_{iA\to jY}(E_i,E_j)}{{\mbox d}E_j},
\label{z-moment}
\end{equation}
with ${\mbox d}n_{iA\to jY}(E_i,E_j)\equiv {\mbox d}\sigma_{iA\to
jY}(E_i,E_j)/\sigma_{iA}(E_i)$. In the decay process, the scattering
length $\lambda_i$ is replaced by the decay length $d_i$, whereas the
${\mbox d}n_{iA\to jY}(E_i,E_j)$ is replaced by the ${\mbox
d}\Gamma_{i\to jY}(E_i,E_j)/\Gamma_i(E_i)$. We note that this part
of the atmospheric $\nu_{\mu}$ flux is isotropic, 
 unlike the contributions
from the $\pi$ and the $K$ decays. This difference is 
 attributable to the
lifetime difference between the charm and the $\pi(K)$ mesons. 
 The various $Z$ moments are calculated using the 
 next-to-leading order perturbative 
 QCD \cite{Nason:1989zy,Mangano:1991jk}.  
The decay moments $Z_{h\nu_{\mu}}$ are calculated by using the
charmed-hadron decay distributions given in
Refs.~\cite{bugaev,Lipari:1993hd}.

\begin{figure}
\epsfig{file=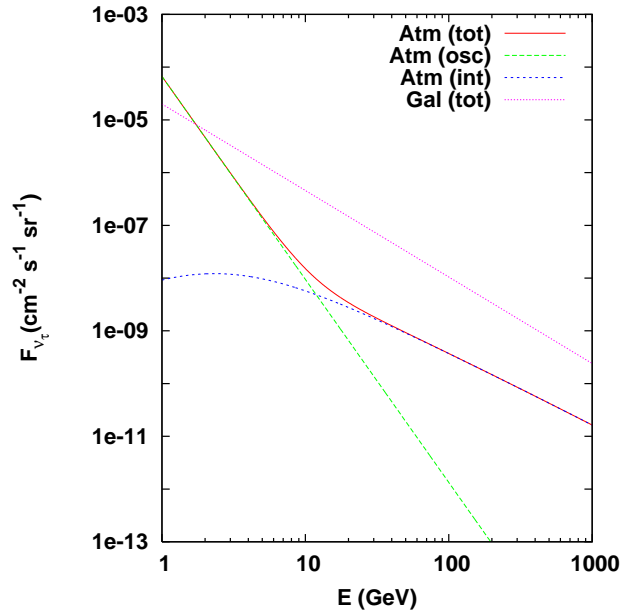,width=3.25in}
\caption{Comparison of the total galactic plane tau neutrino flux with 
 the total downward going atmospheric neutrino flux for the best 
 fit values of the neutrino mixing parameters as a function of the 
 neutrino energy. 
 The intrinsic and oscillated  parts of the total downward going atmospheric 
 tau neutrino flux are also shown as a function of the neutrino energy.}
\label{Fig3}
\end{figure}

We show the relative contributions by the $\pi$, the $K$, 
 and the charm decays
to the overall $\nu_{\mu}$ flux in Fig.~\ref{Fig2}. This is an
extension of the Fig. 3 in Ref.~\cite{Gaisser:2001sd}, where 
 only the $\pi$ and the $K$
contributions are compared. It is obvious 
 that the $\pi$ decay contribution 
dominates for 1 GeV $\leq E \leq 10$ GeV, while the $K$ 
 decay contribution dominates
between $10^{3}$ GeV and $10^{5}$ GeV. The fraction of the charm-decay
contribution rises rapidly at $E \geq 10^{5}$ GeV and becomes dominant
for $E > 10^{6}$ GeV. In this energy range, both the $\pi$ and the $K$ lose
large fractions of their energies before decaying into the neutrinos.

Additionally, the intrinsic atmospheric $\nu_{\tau}$ flux is also
required to completely determine the total atmospheric $\nu_{\tau}$
flux. This flux is calculated using the perturbative QCD, 
 since $\tau$ neutrino arises from the 
 $D_{s}$ decays \cite{Pasquali:1998xf}. The
flux is written as Eq.~(\ref{mu-charm}) with $Z_{ph}$ replaced by the 
$Z_{pD_s}$ and $Z_{h\nu_{\tau}}$ replaced by the $Z_{D_s\nu_{\tau}}$. We
\begin{figure}
\epsfig{file=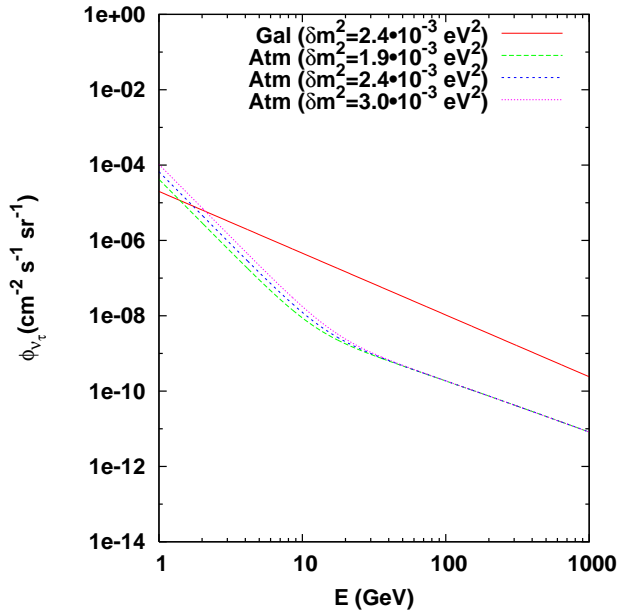,width=3.25in}
\caption{The comparison of the galactic-plane and
the downward going atmospheric $\nu_{\tau}$ fluxes in the presence
of neutrino oscillations with maximal mixing as a function of the 
  neutrino energy in GeV. For
$\delta m^2=2.4\cdot 10^{-3}$ eV$^2$, both fluxes cross at $E=2.3$
GeV.}
\label{fig:down}
\end{figure}
note that the $Z_{D_s\nu_{\tau}}$ contains two contributions. One arises
from the decay $D_s\to \nu_{\tau}\tau$; the other follows from the
subsequent tau-lepton decay, $\tau\to \nu_{\tau}+Y$. The latter
contribution is calculated using the decay distributions of the
decay modes $\tau\to \nu_{\tau}\rho$, $\tau\to \nu_{\tau}\pi$,
$\tau\to \nu_{\tau}a_{1}$ \cite{Li:1995aw,Pasquali:1998xf}, and the $\tau\to
\nu_{\tau}l\nu_{l}$ \cite{Gaisser:1990vg,Lipari:1993hd}
 (Fig. \ref{Fig1} shows the intrinsic downward going atmospheric muon and 
tau neutrino fluxes for 1 GeV $\leq E \leq 10^{3}$ GeV). 

We have calculated the $\phi^{\rm tot}_{\nu_{\tau}}(E)$ 
 by applying Eq.~(\ref{osc}) with
$\phi_{\nu_{\mu,\tau}}(E)$ given by
$\mbox{d}^2N_{\nu_{\mu,\tau}}(E,X)/\mbox{d}E\mbox{d}X$ and
integrating over the slant depth $X$. For $\xi <
70^{\circ}$, the oscillation probability
$P(\nu_{\mu}\to \nu_{\tau})$ is calculated using the relation
$X=X_{0}\exp(-L\cos\xi/h_0)/\cos\xi$ with
  $X_0=1030$ g/cm$^2$ and $h_0=6.4$ km. Here
$L$ is the linear distance from the neutrino production point to the
detector on the Earth \cite{Gaisser:1997eu}. This gives, for instance, 
$P(\nu_{\mu}\to \nu_{\tau})\simeq \sin^{2}(4.5\cdot 10^{-2}({\rm GeV}/E))$
 at $\xi =0^{\circ}$, 
 for the best-fit values of the neutrino mixing parameters.  
\section{The comparison} 

The comparison of the galactic-plane and
the downward going atmospheric $\nu_{\tau}$ flux is given in
 Fig.~\ref{Fig3} and 
Fig.~\ref{fig:down} in the presence of neutrino oscillations
  (we plot ${\rm d}N_{\nu}/{\rm d}({\rm
log}_{10}E)$ instead of ${\rm d}N_{\nu}/{\rm d}E$). The former flux
clearly {\tt dominates} the latter for $E \geq 10$ GeV, whereas the
two fluxes cross at $E =2.3$ GeV for $\delta m^2=2.4\cdot 10^{-3}$
eV$^2$ and $\sin^{2}2\theta =1$. 
 This comparison is however subject to the uncertainty of
galactic-plane $\nu_{\tau}$ flux by the choices of the density $n_p$
and the distance $R$ mentioned before.

One can see that the
atmospheric $\nu_{\tau}$ flux is sensitive to the value of $\delta
m^2$ for $E \leq 20$ GeV. Furthermore, a change of slope occurs for
the atmospheric $\nu_{\tau}$ flux at $E \approx 20$ GeV. Beyond this
\begin{figure}
\epsfig{file=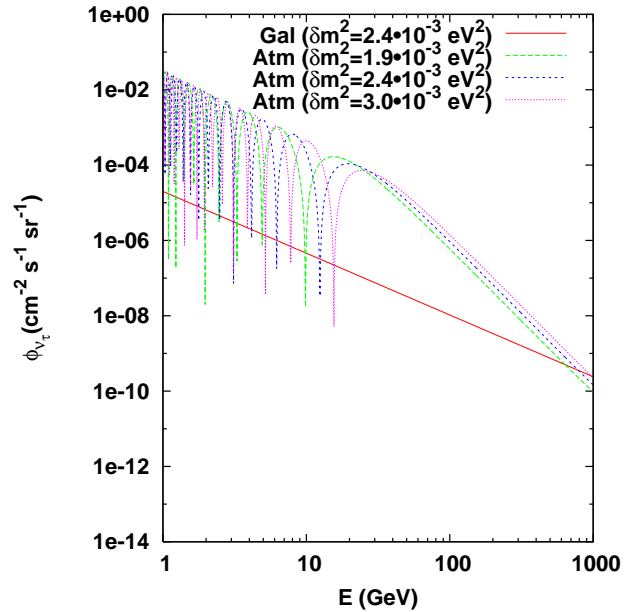,width=3.25in}
\caption{The comparison of the galactic-plane
$\nu_{\tau}$ flux and the atmospheric $\nu_{\tau}$ flux coming in
the zenith angle $\xi=180^{\circ}$ in the presence of neutrino 
 oscillations with maximal mixing as a 
 function of the neutrino energy in GeV.
For $\delta m^2=2.4\cdot 10^{-3}$ eV$^2$, 
 both fluxes cross at $E=800$ GeV.}
\label{fig:60deg}
\end{figure}
energy, the slope of the atmospheric $\nu_{\tau}$ flux is identical to
that of the galactic-plane $\nu_{\tau}$ flux. {\tt For $E > 20$ GeV, the
atmospheric $\nu_{\tau}$ flux is intrinsic}, i.e., coming from the $D_s$
decays, whereas the galactic-plane $\nu_{\tau}$ flux arises from the
oscillation of the $\nu_{\mu}$, which is produced mainly by the $\pi$
decays. In both cases, the hadrons decay before interacting with the
medium. Such a feature dictates the slope of the outgoing neutrino
flux. Below 20 GeV, however, the atmospheric $\nu_{\tau}$ flux
predominantly comes from the $\nu_{\mu}$ oscillations, i.e.,
$\phi^{\rm tot}_{\nu_{\tau}}(E)\approx \phi_{\nu_{\mu}}(E)\cdot
\sin^2 2\theta\cdot \sin^2(L/L_{\rm osc})$ following
Eq.~(\ref{osc}). Since $L_{\rm osc}\equiv 4E/\delta m^2\approx 330$
km for $E=1$ GeV with $\delta m^2=2.4\cdot 10^{-3}$ eV$^2$, we
approximate $\sin^2(L/L_{\rm osc})$ with $(L/L_{\rm osc})^2$ so that
$\phi^{\rm tot}_{\nu_{\tau}}(E)\sim \phi_{\nu_{\mu}}(E)E^{-2}$.
Because the neutrino oscillation effect steepens the
$\phi_{\nu_{\tau}}$ spectrum for $E \leq 20$ GeV, the slope change
of $\phi_{\nu_{\tau}}$ at $E\approx 20$ GeV is significant.
\begin{figure}
\epsfig{file=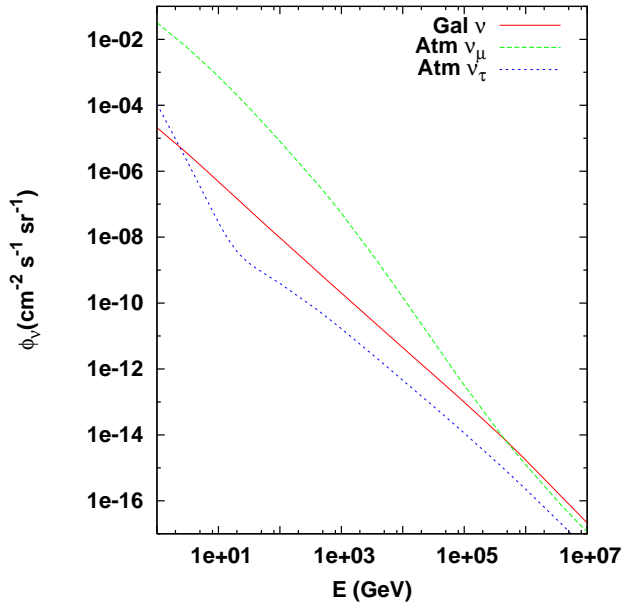,width=3.25in}
\caption{An illustrative comparison of the downward going
atmospheric $\nu_{\mu}$ and $\nu_{\tau}$ fluxes and the
corresponding galactic-plane neutrino fluxes in the presence of
neutrino oscillations as a function of the neutrino energy in 
 GeV. The best fit values of the 
 neutrino mixing parameters, namely, 
 $\delta m^2=2.4\cdot 10^{-3}$ eV$^2$  and $\sin^{2}2\theta =1$, 
 are used. The galactic-plane and the atmospheric
$\nu_{\mu}$ fluxes cross at $E = 5\cdot 10^{5}$ GeV.}
\label{fig:mt}
\end{figure}

We have also worked out the comparison of the galactic-plane and the
atmospheric $\nu_{\tau}$ fluxes for several other zenith angles. 
 For instance, the
crossing of the galactic-plane and the atmospheric $\nu_{\tau}$ fluxes
occurs at $E = 6.0$ GeV for the zenith angle
 $\xi=60^{\circ}$ for $\delta m^2=2.4\cdot 10^{-3}$ eV$^2$ with the
 maximal mixing. Essentially, the atmospheric $\nu_{\tau}$
flux increases with the zenith angle $\xi$ while the galactic-plane
$\nu_{\tau}$ flux stays unchanged.  Moreover,
from the phenomenological point of view, a larger $\xi$ only implies
a larger atmospheric tau neutrino background while our focus is on
the prospective observation of the astrophysical tau neutrinos.

Two key factors determine the
angular behavior of the former flux. First, the atmosphere depth is
larger for $\xi=60^{\circ}$ compared to the downward direction.
Second, the atmospheric muon neutrinos are produced more far away
from the ground detector in this angle, implying a larger
$\nu_{\mu}\to \nu_{\tau}$ oscillation probability. We have found that,
for $E=10$ GeV, the downward going muon neutrinos are produced on
average about $13$ km from the ground detector. At $\xi=60^{\circ}$,
the above distance increases to $34$ km. 
In the angular range $\xi < 70^{\circ}$, the
curvature of the Earth can be neglected and the angular dependence
of the neutrino flux can be easily calculated \cite{Gaisser:1997eu}. 

In Fig.~\ref{fig:60deg}, we show the comparison for the zenith angle
 $\xi=180^{\circ}$ (vertically upward going direction). 
 For $\delta m^2=2.4\cdot 10^{-3}$ eV$^2$ with the
 maximal mixing the crossing between the two fluxes cross at 800 GeV.
 The oscillatory behavior 
in the total atmospheric tau neutrino flux is a manifestation of the 
narrow zenith angle dependence.

We have so far neglected the 3-body muon decay contribution to the
atmospheric $\nu_{\mu}$ flux. As mentioned earlier, this
contribution is relevant only for 1 GeV $\leq E \leq $ 10 GeV. The ratio
of this contribution to the total atmospheric $\nu_{\mu}$ flux is
available in the literature \cite{Lipari:1993hd,Gaisser:1997eu}. 
 For $\cos\xi=0.4 \
(\xi=66^{\circ})$, $E=1$ GeV, the ratio is about $50\%$. It drops to
$30\%$ at $E=10$ GeV for the same $\xi$ \cite{Lipari:1993hd}. This ratio
is not very sensitive to the zenith angle $\xi$. The error due to
neglecting this contribution propagates to the determination of the
atmospheric $\nu_{\tau}$ flux by the neutrino oscillation effect.
Numerically, this error is comparable to the uncertainty of the  
$\nu_{\tau}$ flux due to the uncertainty of the $\delta m^2$, as
$\phi^{\rm tot}_{\nu_{\tau}}(E)\approx \phi_{\nu_{\mu}}(E)\cdot
\sin^2 2\theta\cdot (\delta m^2\cdot L/4E)^2$.

\section{Conclusions} 

The results presented in Figs.~\ref{fig:down}, 
\ref{fig:60deg} and \ref{fig:mt} indicate the opportunities 
 for the tau neutrino
astronomy in the GeV energies for the incident 
 zenith angles $0^{\circ}\leq
\xi \leq 180^{\circ}$ in the two neutrino flavor mixing approximation. 
 The galactic-plane tau neutrino flux {\tt dominates} over
the atmospheric tau neutrino background until a few GeV's.
Furthermore, for $E\leq 20$ GeV, the former flux has a rather
different slope from that of the latter. This is an important
criterion for distinguishing the two fluxes, particularly given that
the normalization of the galactic-plane tau neutrino flux is still
uncertain. 

We have pointed out that the dominance of the galactic-plane tau
neutrino flux over its atmospheric background in GeV energies is
{\tt unique} among all the considered neutrino flavors. 
 This is depicted in
Fig.~\ref{fig:mt}. Because of the $\nu_{\mu} \to \nu_{\tau}$ 
 neutrino oscillations, the {\tt total}
galactic $\nu_{\tau}$ flux is identical to that of the galactic
$\nu_{\mu}$ flux. However, the atmospheric $\nu_{\mu}$ flux is much
 larger than the atmospheric $\nu_{\tau}$ flux. As a result, in
the {\tt presence} of
 neutrino oscillations, the crossing energy value 
for the galactic-plane and the atmospheric $\nu_{\mu}$ 
 fluxes is pushed up
to $5\cdot 10^5$ GeV, which is drastically different from the tau
neutrino case.

We have estimated the galactic-plane tau neutrino induced shower
event rate $N_{\nu_{\tau}}$ for the forthcoming 
 one Mega ton class of detectors. It
is obtained by convolving the total galactic-plane tau neutrino flux
shown in Fig.~\ref{fig:mt} with the charged-current $\nu_{\tau}N$
interaction cross section (using the CTEQ6 parton distribution
functions \cite{Kretzer:2003it}). It is found that the $N_{\nu_{\tau}}$
can be 4.5 (7.5) for $E > $ 10 GeV with a data taking period of 3
(5) years.

A remark concerning the possible background induced by the 
 electron neutrino events and the neutral current events 
 to the prospective observation of tau neutrino events 
 is in order. Below $ 10^{3}$ GeV,
the tau lepton decay length is less than a mm. This tau 
lepton is produced in the detector in the galactic-plane 
tau neutrino induced interactions. 
 There are certain  specific signatures of the
 tau neutrino induced tau leptons such as the appearance
 of the kink at the tau lepton decay 
 (absent for the electrons) \cite{Pessard:2005kv}, 
as well as  the relative
 characteristic fractional energy sharing from the incident 
neutrino \cite{Stanev:1999ki}. 
A large scale finely grained detector with a 
 resolution of a few $\mu $m
will thus be required to disentangle the galactic-plane 
tau neutrino induced events from the events induced by the
 electron neutrinos and the neutral current events 
 on the event-by-event basis.

{\bf Acknowledgements.---}
H.A. thanks Physics Division of NCTS for support.
F.F.L and G.L.L. are supported by the National
Science Council of Taiwan under the grant number
NSC92-2112-M-009-038.

\end{document}